\documentclass[epjCONF]{svjour}
\usepackage{graphics}
\usepackage[varg]{txfonts} 
\usepackage[latin1]{inputenc}
\session-title{International Symposium on Future Directions in UHECR Physics, CERN 2012}
\begin{document}
\title{Radar reflection off extensive air showers}
\author{J. Stasielak\inst{1}\fnmsep\thanks{\email{jaroslaw.stasielak@ifj.edu.pl}} \and S. Baur\inst{2} \and M. Bertaina\inst{3} \and  J. Bl\"{u}mer\inst{2} \and A. Chiavassa\inst{3} \and R. Engel\inst{2} \and A. Haungs\inst{2} \and T. Huege\inst{2} \and K.-H. Kampert\inst{4} \and H. Klages\inst{2} \and M. Kleifges\inst{2} \and O. Kr\"{o}mer\inst{2}\and M. Ludwig\inst{2} \and 
S. Mathys\inst{4} \and P. Neunteufel\inst{2}  \and J. Pekala\inst{1} \and J. Rautenberg\inst{4} \and M. Riegel\inst{2} \and M. Roth\inst{2} \and F. Salamida\inst{5} \and H.  Schieler\inst{2} \and R. \v{S}m\'{\i}da\inst{2} \and M. Unger\inst{2} \and M. Weber\inst{2} \and F. Werner\inst{2} \and H. Wilczy\'{n}ski\inst{1} \and J. Wochele\inst{2} 
}

\institute{Institute of Nuclear Physics PAN, Krakow, Poland 
\and Karlsruhe Institute of Technology (KIT), Karlsruhe, Germany 
\and Universit\`{a} di Torino and Sezione INFN, Torino, Italy
\and Bergische Universit\"{a}t Wuppertal, Wuppertal, Germany
\and Universit\`{a} dell'Aquila and INFN, L'Aquila, Italy
}

\abstract{
We investigate the possibility of detecting extensive air showers by the radar technique. Considering a bistatic radar system and different shower geometries, we simulate reflection of radio waves off the static plasma produced by the shower in the air. Using the Thomson cross-section for radio wave reflection, we obtain the time evolution of the signal received by the antennas. The frequency upshift of the radar echo and the power received are studied to verify the feasibility of the radar detection technique. 
} 
\maketitle
\section{Introduction}
\label{intro}

Traditional techniques of extensive air shower (EAS) detection include counting the shower particles at the ground level and measuring fluorescence light from the excited nitrogen in the atmosphere. Detecting radio emission from the shower particles is a promising new technique that is currently in development. An alternative method is the radar technique, where a ground-based radio transmitter illuminates the ionization trail left behind the shower front and another ground-based antenna receives the scattered radio signal with its upshifted frequency. 

The concept of implementing a radar for cosmic ray detection was first introduced about 70 years ago by Blackett and Lovell \cite{blackett}. However due to the negative results of dedicated experiments, this concept was forgotten for several decades. 
In recent years renewed attention has been attracted to this topic, see e.g. \cite{gorham,gorham2,bakunov,takai}. 
Experimental efforts to detect EAS using the radar technique were
made by several groups, e.g. \cite{matano}, the LAAS group \cite{lyono}, \cite{terasova}, the MARIACHI experiment \cite{mariachi}, or the TARA project \cite{tara}, however no detection was confirmed so far.

In this paper we investigate the possibility of detecting EAS by the bistatic radar system. Especially, we are interested in verifying the feasibility of supplementing the CROME (Cosmic-Ray Observation via Microwave Emission) detector \cite{smida,smida2} with a radio transmitter and using its existing antennas as receivers for the radar echo. 
The CROME antenna setup consists of several microwave receivers
for the frequency ranges $1.2-1.7$ (L band), $3.4-4.2$ (extended C band), and
$10.7-11.7$ GHz (low Ku band). Since we expect large frequency upshifts of the radar echo and the CROME antennas are fine tuned to the GHz frequency band, we can narrow down our interest to the frequency range of 
$1-100$ MHz of the emitted radio waves.

The concept of detecting EAS using radar technique is based on the principle of scattering radio waves off the static plasma produced in the atmosphere by the energetic particles of the shower. The locally produced plasma decays exponentially with the lifetime $\tau$. According to Vidmar \cite{vidmar}, for the plasma densities relevant for EAS and at low altitudes, the three-body attachment to oxygen dominates the
deionization process as it depends quadratically on the oxygen density. This leads to the plasma lifetime of 10 ns at sea level and about 100 ns at altitude of 10 km. Since the received power of the radar echo is strongly diminished by the geometrical factor, the strongest signal will be obtained from the altitudes close to the ground level. Thus we can safely assume $\tau=10$ ns.

The ionization trail that results from meteor or lightning is traditionally divided  into the underdense and overdense regions, depending on the local plasma frequency $\nu_p$. If the electron density is high enough that the plasma frequency exceeds the radar frequency, then the region is  overdense and the radio wave is reflected from its interface. In contrast, if the local plasma frequency is lower than the frequency of the incoming radio wave, then the region is underdense and the radio wave can penetrate the ionized region. In such a case the reflections are caused by the Thomson scattering of the radio wave on individual free electrons.

If we consider the radar frequency in the range $1-100$ MHz and shower energies up to $10^{20}$ eV, then the maximum size of the overdense plasma region will be several meters. Now, if we calculate the attenuation length of the radio wave in the plasma with frequency $\nu_p = 100$ MHz and $\nu_p = 1$ MHz
one obtains $\sim$300 m and $\sim$3000 km, respectively. Those lengths are much larger than the size of the overdense plasma, thus the radio wave can easily penetrate the whole volume of the plasma disk produced by the EAS.

Using the Thomson cross-section for radio wave reflection seems to be justified from the above arguments. 
An exact calculation of the total power received requires the integration of the contributions from each individual electron in the shower. 
The final result will then depend on the individual phase factors of the scattering electrons.
 
\section{Radar reflection}
\label{sec:1}

\begin{figure}
\begin{center}
\resizebox{0.65\columnwidth}{!}{%
  \includegraphics{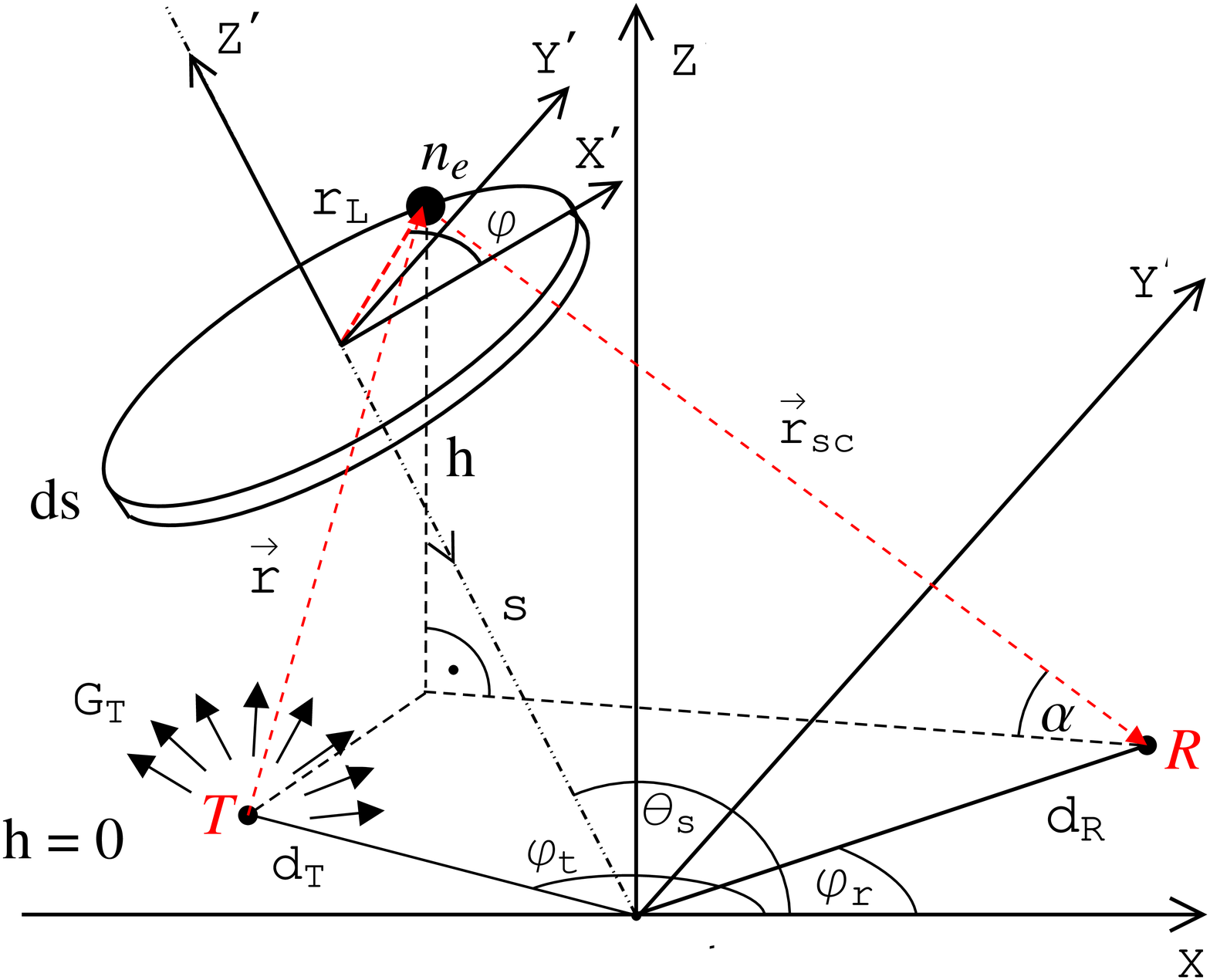} }
\caption{A schematic diagram representing a bistatic radar system and reflection from the static plasma disk produced by the EAS in the atmosphere. See the text for a detailed explanation.}
\label{fig:1} 
\end{center}
\end{figure}

\begin{figure}
\begin{center}
\resizebox{0.7\columnwidth}{!}{%
  \includegraphics{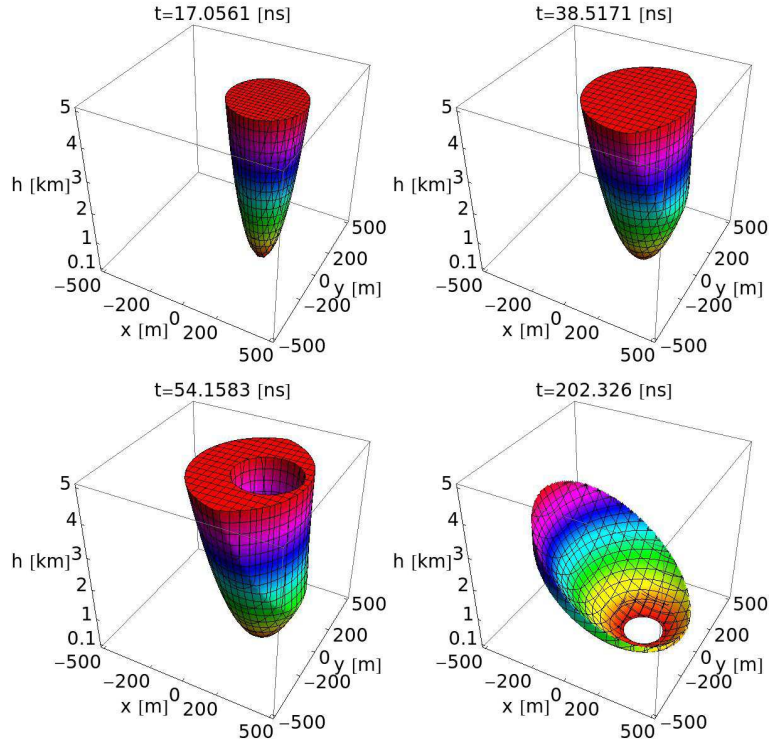} }
\caption{The time evolution of the region from which the scattered radio waves arrive simultaneously to the detector. The arrival time
$t=0$ corresponds to the moment at which the shower hits the ground.
The considered shower has energy of $10^{18}$ eV and is heading towards the transmitter (x=0 m, y=0 m).
The receiver (x=200 m, y=0 m) is placed 200 m away from the transmitter. The colors scale with the altitude.}
\label{fig:boundary} 
\end{center}
\end{figure}

A schematic diagram representing a bistatic radar system is shown in Figure \ref{fig:1}.
A ground-based radio transmitter (T) irradiates a disk-like shower. The front edge of the disk is an ionization front moving towards the ground with the speed of light and leaving created static plasma behind it. The radio signal is scattered by the free electrons in the ionization trail and subsequently received by the ground-based antenna (R). The geometry of the bistatic radar system is set up by the polar coordinates of the transmitter and the receiver, i.e. by the distances from the shower core to the transmitter ($d_T$) and to the receiver ($d_R$) together with the angles $\varphi_t$ and $\varphi_r$. 

Let us consider the radar reflection from the disk element with coordinates ($r_L$, $\varphi$), altitude h, and electron density $n_e$. Its contribution to the radar echo at the receiver at time t can be expressed by 
\begin{equation}
	U_{rcv} (t,s,r_L,\varphi) =U_T \sqrt{G_T}
 e^{i\omega t + \varphi_0} 
{e^{-i\int_{{\bf r}} n{\bf k} \cdot d{\bf r}} e^{ - i\int_{{\bf r_{sc}}} n{\bf k_{sc}} \cdot d {\bf r_{sc}} } \over |{\bf r}| |{\bf r_{sc}}|} \sin \alpha \sqrt{   \frac{d \sigma_T}{ d \Omega}}  n_e 
\rm{,} \label{dUrcv}
\end{equation}
where $G_T$ is the transmitter antenna gain, $U_T$ [V] is the magnitude of the transmitted field, $\bf k$ and $\bf k_{sc}$  are the wave vectors of incoming and scattered radio wave, $\varphi_0$ is the initial phase of the emitted signal, $d\sigma_T/d\Omega$ is the Thomson cross-section, s is the projection of the distance between the shower core and the considered disk element on the shower axis, $\alpha$ is the inclination angle of the reflected radio wave, and n is the refractive index of the air. By analogy to the setup of the CROME detector, we assume that the receiver is oriented vertically upwards. 

The signal received by the antenna at a given time t is sum of the signals scattered at different times and from different parts of the plasma disk. These individual contributions interfere with each other and only an integral over the whole volume $vol(t)$ from which they arrive simultaneously gives us the correct value. The total electric field strength of the radio wave at the receiver is given by
\begin{equation}
	U_{rcv}(t) = \int_{vol(t)} U_{rcv} (t,s,r_L,\varphi) r_L dr_L d\varphi ds  \rm{.} \label{total}
\end{equation}
Note that time $t$ is defined in such a way that $t=0$ coincides with the moment at which the shower hits the ground.

Figure \ref{fig:boundary} shows the time evolution of the region of integration $vol(t)$ 
(see equation (\ref{total})) for a shower with an energy of $10^{18}$ eV heading towards transmitter (x=0 m, y=0 m). The receiver (x=200 m, y=0 m) is placed 200 m away from the transmitter. The colors scale with the altitude.

As we can see, the volume over which we integrate the signal spans usually over the wide range of altitudes. This implies a high variation of the air density. Therefore, the refractive index and the Cherenkov angle can vary significantly over the considered region. This makes the whole analysis very complex.

\section{Frequency upshift}
\label{sec:2}
\begin{figure}
\begin{center}
\resizebox{0.9\columnwidth}{!}{%
  \includegraphics{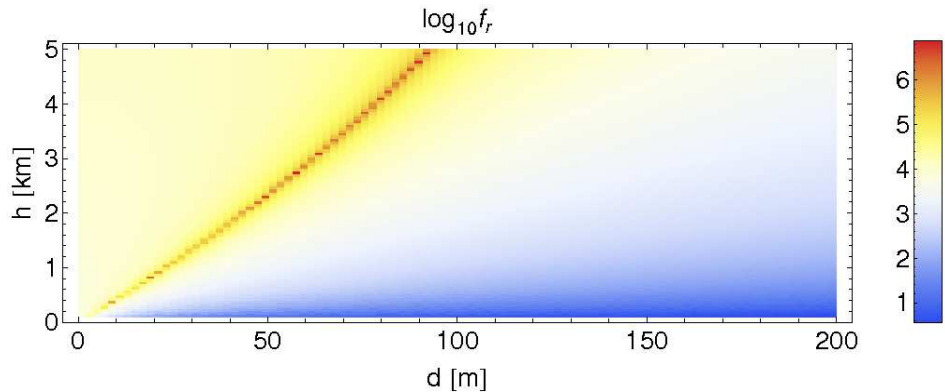} }
\caption{Upshifts $f_r$ of the radio wave scattered on different parts of the shower disk, which is heading vertically towards the transmitter. The altitude of the disk element is given by $h$, whereas its distance to the receiver in the horizontal plane is $d$. }
\label{fig:upshift} 
\end{center}     
\end{figure}

Despite the fact that the radio wave is scattered on the static plasma, the region in which the plasma is created moves with relativistic velocity. Therefore, we can observe the Doppler shift of the received radar echo. Figure \ref{fig:upshift} shows the upshifts $f_r$ of the radio wave scattered on different parts of the shower disk, which is heading vertically towards the transmitter. The altitude of the disk element is given by $h$, whereas its distance to the receiver in the horizontal plane is $d$. 

The frequency upshift depends on the wave direction and the refractive index of air. It has the highest value for the case in which the viewing angle coincides with the Cherenkov cone. 
The typical $f_r$ is high enough to upshift the MHz signal into the GHz range. Note that $f_r$ only weakly depends on the distance from the shower core to the transmitter $d_T$.

\section{Details of the simulation}
\label{sec:3}
The main assumptions and approximations are the following:
\begin{itemize}
	\item The shower longitudinal profile is taken to be a fit to the Gaisser-Hillas function for proton showers.
	\item The lateral distribution of the shower is given by the Gora function \cite{gora}.
	\item Static plasma produced by the shower decays exponentially with the characteristic time $\tau=10$ ns.
	\item We consider radar reflections only from the part of the ionization trail which is up to 40 ns behind the shower front. In other words, the length of the plasma disk produced by the extensive air shower is constrained to $4\tau$. This leads to less than 2$\%$ of the total plasma electrons not being accounted for.
	\item We use the US Standard Atmosphere model.
	\item We assume that the initial phase of the emitted signal is $\varphi_0=0$. Choosing different values of $\varphi_0$ can lead to slightly different values of the received power.
	\item The effective area of the receiver antenna is $A_R$=1 m$^2$ and the transmitter emits signal over the whole upper hemisphere ($G_T=2$).
\item The receiver points vertically and its sensitivity is independent of the direction and frequency of the radar echo. The detector efficiency is 100$\%$. 	
\end{itemize}

\section{Power received by the detector antenna}
\label{sec:4}
The ratio of the 'instantaneous' power $P_R(t)$ received by the detector antenna to the power emitted by the transmitter $P_T$ is given by
\begin{equation}
	P_R(t)/P_T = R^2(t)
\end{equation}
where
\begin{equation}
	R(t) = \frac{1}{\sqrt{4\pi}}\frac{\rm{Re} U_{rcv} (t)}{U_T}\sqrt{A_R}
\end{equation}
The term $R(t)$ is proportional to the electric field strength detected by the receiver. It can be used in the Fourier analysis to obtain the power spectrum of the recorded signal. The 'real' power received by the detector $P_R$ can be obtained by averaging $R^2(t)$ (according to the time resolution of the detector). The ratio $P_R/P_T \sim G_T A_R$.

\begin{figure}
\begin{center}
\resizebox{0.89\columnwidth}{!}{%
  \includegraphics{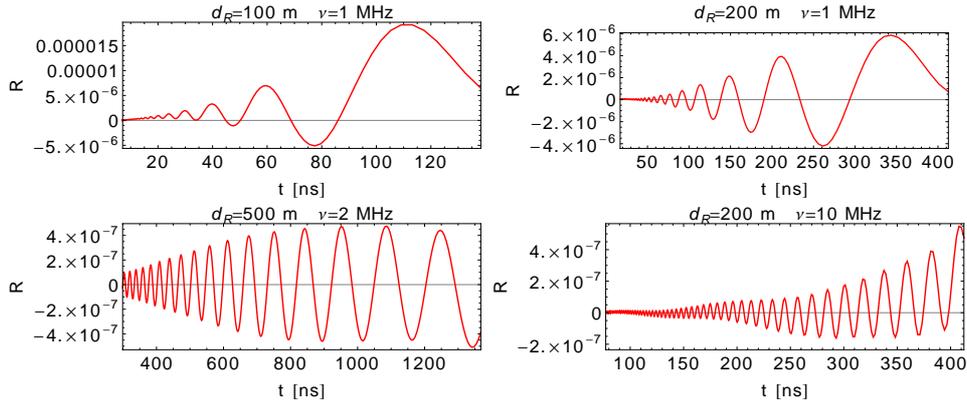} }
\caption{The waveforms R(t) of the radar echoes for different frequencies $\nu$ of the emitted radio wave and different shower core-receiver distances $d_R$. In all cases a vertical shower with energy $10^{18}$ eV heading towards the transmitter is considered.}
\label{waveR}   
\end{center}
\end{figure}

\section{Results}
\label{sec:5}
Figure \ref{waveR} shows the waveforms R(t) of the radar echoes for different frequencies $\nu$ of the emitted radio wave and different shower core-receiver distances $d_R$. In all cases the receiver is outside the Cherenkov cone (which is the most common case) and vertical showers heading towards the transmitter are considered. The time $t=0$ coincides with the moment at which the shower hits the ground. 

As we can see, the frequency of the received signal is higher than that of the emitted one. Moreover, the amplitude grows with time,
which is the geometrical effect of the reflections from the lower parts of the atmosphere.


For showers, where the receiver is inside the Cherenkov cone, the time sequence is reversed and the lower part of the shower is seen first. Therefore, the amplitude decreases with time. However, this is the case only for very inclined showers with cores up to several meters away from the receiver.  

The ratios of the power received by the detector to the emitted one ($P_R/P_T$) for vertical showers are given in table \ref{taba}. It is evident that the strength of the signal decreases with the frequency $\nu$. It is simply because the size of the region from which we get the coherent signal decreases with the wavelength and the negative interference cancels out the signal.

\section{Conclusions}
\label{sec:6}
The typical ratio of the received to the emitted power is in the range of $10^{-10}$ - $10^{-13}$, which makes the detection of the radar reflection plausible. The C band ($\sim$3 GHz) is particularly well-suited for this purpose because of very low noise ($< 10$ K) in this frequency range. The CROME antennas supplemented with commercial high power MHz transmitter would create a radar system dedicated to EAS detection.
A note should be added, that the shown time traces are for an infinite bandwidth detector $-$ a realistic detector would only be able to detect the signal in narrow frequency range. Moreover, it will only see the shower for a certain fraction of its development.

The performed analysis will help to choose the frequency of the emitted wave to optimize the detection of the radar echo. The work is in progress.

\begin{table}
\caption{The ratios of the power received by the detector to the emitted one ($P_R/P_T$) in dB for different frequencies $\nu$ of the emitted radio wave, different transmitter-receiver distances $d_R$ and different shower energy $E$. In all cases a vertical shower heading towards the transmitter is considered.}
\label{taba}      
\begin{center} 
\begin{tabular}{lllll}
\hline\noalign{\smallskip}
$E=10^{18}$ eV  \\
$\nu$ \textbackslash $d_R$ & 50 m & 100 m & 200 m & 500 m \\
\noalign{\smallskip}\hline\noalign{\smallskip}
1 MHz & -93 dB & -101 dB & -111 dB & -126 dB \\
2 MHz & -101 dB & -108 dB & -117 dB & -131 dB \\
5 MHz & -110 dB & -117 dB & -129 dB & -139 dB \\
10 MHz & -117 dB & -127 dB & -137 dB & -150 dB \\
\noalign{\smallskip}\hline
\end{tabular}
\begin{tabular}{llll}
\hline\noalign{\smallskip}
$d_R=200$ m  \\
$\nu=1$ MHz \\
$E$ [eV] & $10^{18}$ eV & $10^{19}$ eV & $10^{20}$ eV \\
\noalign{\smallskip}\hline\noalign{\smallskip}
& -111 dB & -88 dB & -67 dB  \\
\noalign{\smallskip}\hline
\end{tabular}
\end{center}
\end{table}

\section{Acknowledgments}
It is our pleasure to acknowledge the interaction and collaboration
 with many colleagues from the KASCADE-Grande and Pierre Auger Collaboration.
This work has been supported in part by the KIT start-up grant 2066995641,
 the ASPERA project BMBF 05A11VKA, the Helmholtz-University Young
 Investigators
 Group VH-NG-413 and the National Centre for Research and Development
 (NCBiR).

\end{document}